\documentclass[aps,pre,twocolumn,showpacs]{revtex4}
\usepackage{epsfig}
\usepackage{amsmath}
\begin{document}

\title{Estimating the value of containment strategies in delaying the arrival time of an influenza pandemic:
A case study of travel restriction and patient isolation}

\author{Lin Wang, Yan Zhang, Tianyi Huang, and Xiang Li\footnote{lix@fudan.edu.cn}}

\affiliation
{Adaptive Networks and Control Laboratory, Department of Electronic Engineering, Fudan University, Shanghai 200433, P.R.China}

\begin{abstract}
With a simple phenomenological metapopulation model, which characterizes the invasion process of an influenza pandemic from
a source to a subpopulation at risk, we compare the efficiency of inter- and intra-population interventions in delaying the
arrival of an influenza pandemic. We take travel restriction and patient isolation as examples, since in reality they
are typical control measures implemented at the inter- and intra-population levels, respectively. We find that the
intra-population interventions, e.g., patient isolation, perform better than the inter-population strategies such as travel
restriction if the response time is small. However, intra-population strategies are sensitive to the increase of the response
time, which might be inevitable due to socioeconomic reasons in practice and will largely discount the efficiency.
\end{abstract}

\maketitle

\section{Introduction}

\vspace{-0.25cm}

During the past decades, extensive efforts have been made to investigate the spread of epidemics. Besides various epidemiological models
having been proposed to explore virus transmission in a closed population\cite{ANDERSONMAY}, the study of network spreading uses
structured populations to understand the evolution of epidemics in more realistic social settings\cite{PR424175,BBV08,RMP801275,NP832}.
These studies have contributed a great deal of insightful findings, such as the absence of epidemic threshold in scale-free networks\cite{PRE63066117}, the reaction-diffusion process, and metapopulation\cite{MATHBIO753,PRE84041936}, to name a few. These
significant advances have raised new issues on how to limit or control the spread of infectious diseases in human society.

To curb the spatial spread of diseases from city to city, a variety of strategies are recommended according to World Health
Organization(WHO) or United States(US) response plans\cite{WHO}: (i) Vaccination of prior groups or dynamic mass vaccination;
(ii) antiviral drugs for prophylaxis and treatment; (iii) community-based prevention and control; and (iv) travel-related containment
measures. Except for the fact that travel-related measures are implemented at the inter-city level, other strategies are mainly
performed at the intra-city level. The first two pharmaceutical interventions cut down the number of potential susceptibles or
allay the virus transmission rate, respectively. Community-based strategies might affect individuals (e.g., patient isolation,
self-isolation, quarantine), groups, or entire communities (e.g., cancellation of public gatherings, school closures) in a city.
Travel-related measures mainly result in the restriction or cancellation of nonessential trips.

By supposing that the outbreak of a pandemic is underway, many works have studied the efficiency of strategies by using the metapopulation
model, which harnesses the reaction-diffusion framework to sketch human daily contacts and mobility. The epidemic reaction takes
place inside each subpopulation due to personal contacts, and the infectious disease cascades subpopulation by subpopulation via
the travel of individuals (here each city is represented by a subpopulation). The importance of various strategies in decreasing the
attack rate or prevalence has been extensively studied in Refs. \cite{SCI326729,SCI3091083} mainly by computational simulations.
Particularly, by analyzing the delay of arrival time of the disease \cite{PLOSMED3e212,NatMed12497,PLoSONE2e401,PLoSONE6e16591,MB21470},
it has been shown that the efficiency of travel restriction in slowing down the international spread of pandemic influenza is limited.

In these seminal works, the intra- and inter-population interventions are seldom compared with each other to provide a holistic picture
about their value in delaying disease invasion. This should give us pause for thought. Whether it is reasonable to discard the tactic
of travel restriction might also depend on how good the intra-population strategies perform. In an attempt to study this issue, we
theoretically analyze the efficiency of two kinds of typical containment strategies, namely, travel restriction and patient isolation,
which are implemented at the inter- and intra-population levels, respectively. We mainly use a simple phenomenological model following Refs. \cite{MB21470,PLoSONE2e143}, which considers the importation of an infectious disease from a source to a region at risk during the early
stage of a pandemic outbreak. Since the spreading process cascades subpopulation by subpopulation, this two-subpopulation
version\cite{PLoSONE6e16591} is a simple model but rational approximation of the initial stage of the pandemic. We mainly focus on the
impact of strategies to delay the arrival time of disease in the subpopulation at risk, because no outbreak will occur in an unaffected
region before the introduction of infectious seeds. After the disease lands in the subpopulation, the ongoing endogenous transmission
will become the mainstream of infections\cite{SCI3091083,PLoSONE6e16591}. Thus the first arrival time of infectious travelers is an
important quantity characterizing the timing of the disease outbreak\cite{MB21470,JSM09001,JTB251509}.

\vspace{-0.25cm}

\section{Model description}

\vspace{-0.25cm}

To build the model, we first specify the mechanism of individual mobility between subpopulations $x,y$. Following Refs. \cite{MB21470,PLoSONE2e143,JSM09001,JTB251509}, at every time step, each individual may travel from his current
location $x$($y$) to a neighboring subpopulation $y$($x$) with a per capita diffusion rate $\omega_{xy}$($\omega_{yx}$).
We define the unit time as 1 day. The model proceeds with discrete time steps. In reality, the amount of transportation
flows, e.g., air traffic, between cities is often symmetric\cite{JSM09001,JTB251509,PNAS1068847}, which indicates a
detailed balance for the traffic flows. For simplicity, we assume that the subpopulations $x,y$ have the same population
size $N_{x}=N_{y}=N$ and diffusion rate $\omega_{xy}=\omega_{yx}=\omega$. Thus there are on average $\omega N$ individuals
departing from each subpopulation per day. Note that relaxing these two restrictions does not change the main results of
this Brief Report as long as we maintain a detailed balance condition. While mobility couples different locations, the epidemic
reaction process occurs in each subpopulation, where the population is mixing homogeneously. We consider a standard
susceptible-infective-removed (SIR) compartment model to represent the influenza-like illness\cite{MATHBIO753,PRE84041936,SCI326729}.
At a given time $t$, the number of susceptible, infectious, and recovered individuals in $x(y)$ are defined as $S_x(t),I_x(t),R_x(t)(S_y(t),I_y(t),R_y(t))$, respectively. The SIR reaction is governed by the transition rates $\mu$ and
$\beta$\cite{ANDERSONMAY}. In a unit time, an infectious one recovers and becomes immune at the rate $\mu$. The parameter
$\beta$ characterizes disease transmissibility, which reflects the combined factors of the virus transmission rate and
individual contact rate per unit time\cite{PRE84041936}. A susceptible individual might acquire infection by contact with
infectious ones staying in the same subpopulation. With the mean-field approximation, at time $t$, the probability for a
susceptible one in subpopulation $x(y)$ to acquire infection is found by multiplying the density of infectious $I_{x}(t)/N(I_y(t)/N)$
by $\beta$\cite{ANDERSONMAY}. In this baseline case, the transfer of susceptible and infectious individuals is ruled by the
diffusion rate $\omega$. The epidemic threshold is determined by the basic reproductive number $R_0=\beta/\mu$, which
identifies the expected number of secondary infections produced by an infected individual during his infectious period in an
entire susceptible population\cite{ANDERSONMAY}.

We next specify the dynamics under interventions. Since many socioeconomic factors might defer the implementation of strategies, we
define a response time $t_0$ representing the time interval between the actual inception of an outbreak and the time when the
strategies become available. Travel restriction (TR) mainly affects individual mobility between two subpopulations. We
define the parameter $\alpha$ as the intensity of TR, which means that a reduction of fraction $\alpha$ in travel begins at
time $t_0$, i.e., in the model, we decrease the diffusion rate from $\omega$ to $(1-\alpha)\omega$ after time $t_0$.

Patient isolation (PI) mainly impacts individual compartment transitions. The effect of PI may relate to enforcement by local
authorities, or is attributed to the self-isolation of infected individuals. For simplicity, we do not distinguish between these two
aspects. The parameter $\eta$ is defined to reflect the intensity of PI. It means that on average a fraction $\eta$ of infectious
persons will be isolated per unit time after $t_0$. We introduce the PI by adding an isolation process that each infectious one has
a likelihood to be isolated with rate $\eta$ per unit time. Since these isolating individuals have little chance to cause infection,
we remove them as long as they are isolated.

\vspace{-0.25cm}

\section{ANALYTICAL AND SIMULATION RESULTS}

\vspace{-0.25cm}

Initially, an infectious individual is introduced into subpopulation $x$. Thus the initial condition is $I_x(0)$=1,$I_y(0)$=0.
We first analyze the efficiency of TR in slowing down disease invasion to subpopulation $y$. The key issue is to evaluate its
impact on delaying the first arrival time(FAT) of infectious travelers from $x$. With the Poisson process
assumption that the diffusion of any individual is independent from that of others, the probability that the first infectious
individual arrives in subpopulation $y$ at time $t^{y}=t$ is
\vspace{-0.25cm}
\begin{equation}
P(t^{y}=t)=[1-(1-\omega)^{I_{x}(t)}]\prod^{t-1}_{t_{i}=1}(1-\omega)^{I_{x}(t_{i})},\label{eq.1}
\end{equation}
which describes that at least one successful transfer of infectious individuals from subpopulation $x$ to $y$ occurs at time $t$,
and none at previous time steps\cite{JSM09001,JTB251509}. In reality, it is general that the number of travelers per day is
several orders of magnitude smaller than the total population of a city, where only small amounts of people leave to travel per
day. Empirical evidence of worldwide or US domestic air transportation\cite{MATHBIO753} suggests that the daily diffusion
rate of individuals on each flight route is of the order $10^{-4}$ or less. We here assume $\omega=10^{-4}$,$N=10^6$. Using
the Taylor expansion, Eq. (\ref{eq.1}) becomes $P(t^{y}=t)=\omega I_{x}(t)\exp[-\omega\sum_{0<t_{i}<t}I_{x}(t_{i})]$.

Based on many seminal works\cite{SCI326729,SCI3091083,PLOSMED3e212,NatMed12497,PLoSONE2e401,PLoSONE6e16591,MB21470}, we assume a
pandemic influenza with $R_0=1.75$ and the infectious period $\mu^{-1}=3$ days. In this case, the Malthusian parameter $\lambda$,
the real-time exponential growth rate at the early stage of an outbreak\cite{PRL103038702,Interface7873}, is $\beta-\mu=0.25$.
Since $\omega\!\ll\!\lambda$, the SIR reaction happens at a time scale much faster than the diffusion process, thus the number of
infectious individuals in subpopulation $x$ grows sufficiently before subpopulation $y$ is invaded. Meanwhile, at this early
stage, the infectious ones only make up a small fraction of the total population in $x, I_x(t)\!\ll\!N$. With a mean-field approximation
for the evolution of infectious individuals, we have\cite{BBV08,JSM09001,JTB251509} $I_{x}(t_{i})\!\simeq\!I_{x}(0)\exp(\lambda t_{i})$, $t_{i}\!\le\!t^{y}$. Using the continuum approximation $\sum_{0<t_{i}<t}I_{x}(t_{i})\!=\!\int^{t}_{0}d\tau I_{x}(\tau)$, we obtain
the probability density of FAT, $P(t)\!=\!\omega\exp[\lambda t-(\omega/\lambda)\exp(\lambda t)]$, with the mean value
$<\!t^{F}\!>=(1/\lambda)(\ln(\lambda/\omega)-\gamma)$\cite{JSM09001,JTB251509}, where $\gamma$ is the Euler constant.
With the above given parameters, this characteristic time scale of FAT is $<\!t^F\!>\simeq29$ days.

\begin{figure*}
\setlength{\abovecaptionskip}{-5pt}
\begin{center}
\includegraphics[width=5in]{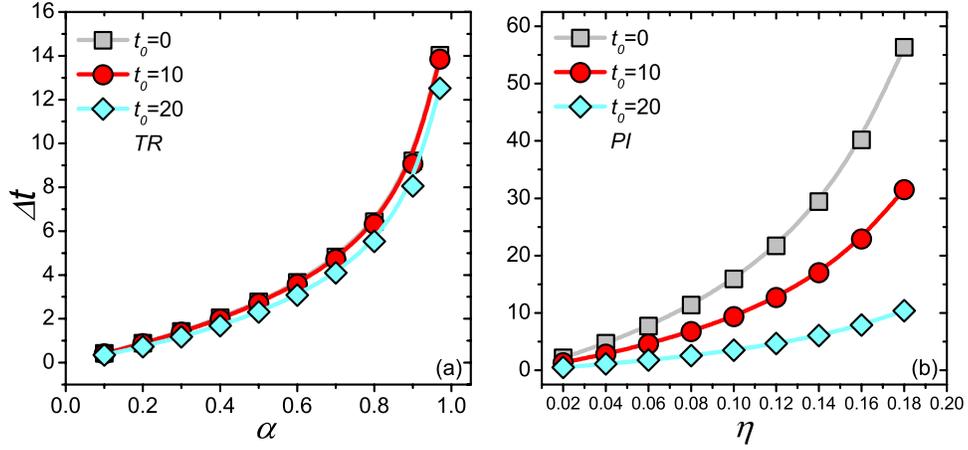}
\end{center}
\caption{(\emph{Color online}) The analytical results of the relation between the delay of FAT, $\Delta t$, and the
intensity of strategies. (a) Travel restriction. (b) Patient isolation. The colored squares, circles, and
diamonds refer to the cases of $t_{0}$=0, 10, and 20 days, respectively.}
\label{fig.1}
\end{figure*}

In the TR scenario, when the FAT is smaller than the response time $t_0$, the probability density of FAT is still $P(t)$; however,
when the FAT is larger than $t_0$, this probability density becomes
$P_{\alpha}(t)=\big{(}1-(1-(1-\alpha)\omega)^{I_{x}(t)}\big{)}\prod_{0<t_{i}<t_{0}}(1-\omega)^{I_{x}(t_{i})}\prod_{t_{0}\le t_{j}<t}\big{(}1-(1-\alpha)\omega)^{I_{x}(t_{j})}\simeq(1-\alpha)\omega\exp[\lambda t-(1-\alpha)\omega\exp(\lambda t)/\lambda
-\alpha\omega\exp(\lambda t_{0})/\lambda]$. We numerically calculate the average FAT through
$<\!t_{\alpha}^{F}\!>=\!\int_0^{t_0}\tau P(\tau)d\tau+\int_{t_{0}}^{\infty}\tau P_\alpha(\tau)d\tau$, and get the delay of FAT,
$\Delta t(\alpha)$, by solving
\begin{equation}
\Delta t(\alpha)=<\!t^{F}_{\alpha}\!>-<\!t^{F}\!>.\label{eq.2}
\end{equation}
If the response time $t_0$ is negligible ($t_0\!=\!0$), Eq.(\ref{eq.2}) is
simplified as
\begin{equation}
\Delta t(\alpha)|_{t_{0}=0}\!=\!-\ln(1-\alpha)/\lambda,\label{eq.3}
\end{equation}
which recovers the results obtained by the cumulative probability $P(t^{y}\leq t)$ in Refs. \cite{PLoSONE6e16591,MB21470}.
Note that Eq (\ref{eq.3}) is independent from the values of $\omega,N$. With $\lambda$=0.25,
unless the intensity $\alpha$ is increased to an unpractically high level ($\alpha>0.97$), $\Delta t(\alpha)$ cannot
be longer than 2 weeks.

To study the PI scenario, we first consider the case where the FAT is larger than $t_0$. At this early stage, we still have the
approximation $I_x(t_{i})\!\simeq\!\exp(\lambda t_{i})$ when time $t_i\!\le\!t_0$; after $t_0$, the Malthusian parameter becomes $\lambda_{\eta}\!=\!\lambda-\eta$, and thus we have $I'_x(t_j)\!\simeq\!\exp(\eta t_0)\exp(\lambda_{\eta}t_{j})$ when
$t_{0}\!<\!t_{j}\!\le\!t^y$. The probability density in this case is
$P_{\eta}(t)=(1-(1-\omega)^{I'_{x}(t)})\prod_{0<t_{i}\le t_{0}}(1-\omega)^{I_{x}(t_{i})}\prod_{t_{0}<t_{j}<t}(1-\omega)^{I'_{x}(t_{j})}\simeq\omega I'_{x}(t)\exp(-\omega\int^{t_{0}}_{0}I_{x}(\tau)d\tau)\exp(-\omega\int_{t_{0}}^{t}I'_{x}(\tau)d\tau)
=\omega\exp[\Theta(t_{0})]\exp[\lambda_{\eta} t\!-\!\omega\exp(\eta t_{0}\!+\!\lambda_{\eta}t)/\lambda_{\eta}]$, where $\Theta(t_{0})\!
=\!\eta t_{0}\!-\!\omega\exp(\lambda t_{0})/\lambda\!+\!\omega\exp(\lambda t_{0})/\lambda_{\eta}$. If the response time is
negligible ($t_{0}$=0), we simplify the former expression as $P_{\eta}(t)|_{t0=0}\!\simeq\!\omega\exp[\lambda_{\eta}t\!-\!\omega \exp(\lambda_{\eta}t)/\lambda_{\eta}]$, which leads to the average FAT, $<t_{\eta}^{F}>|_{t_{0}=0}\!=\!\int_{0}^{\infty}\tau P_{\eta}(\tau)|_{t0=0}d\tau\!\simeq\!(\ln(\lambda_{\eta}/\omega)-\gamma)/\lambda_{\eta}$. In this case, we get the relation between
$\Delta t$ and $\eta$ by solving the equation
\begin{equation}
\Delta t(\eta)|_{t_{0}=0}=<\!t^{F}_{\eta}\!>|_{t_{0}=0}\ -<\!t^{F}\!>\label{eq.4}
\end{equation}
If $t_0>0$, the average FAT is numerically integrated via the equation
$<\!t_{\eta}^{F}\!>=\int_0^{t_0}\tau P(\tau) d\tau+\int_{t_{0}}^{\infty}\tau P_{\eta}(\tau)d\tau$.
We therefore have the relation between $\Delta t$ and $\eta$ as
\begin{equation}
\Delta t(\eta)=<\!t_{\eta}^{F}\!>-<\!t^F\!>.\label{eq.5}
\end{equation}

With Eq. (\ref{eq.4}) and $\lambda=0.25$, we find that an intermediate level of the strategy intensity $\eta=0.12$ can adequately
suspend the arrival of disease to subpopulation $y$ for more than 3 weeks. When the response time $t_{0}=0$, we conclude that the
strategy of PI performs better than the TR. This is mainly because the TR alone can not mitigate the initial exponential growth of
infectious ones in the source. However, the strategy of PI is highly sensitive to the increase of the response time $t_{0}$. As
shown in Fig. \ref{fig.1}, when $t_0$ increases from 0 to 20 days, there is an evident decline for the delay $\Delta t(\eta)$ in
the PI scenario, while the delay $\Delta t(\alpha)$ actualized by implementing the TR is robust to the increase of $t_{0}$.

\begin{figure}[!ht]
\setlength{\abovecaptionskip}{-5pt}
\begin{center}
\includegraphics[width=2.5in]{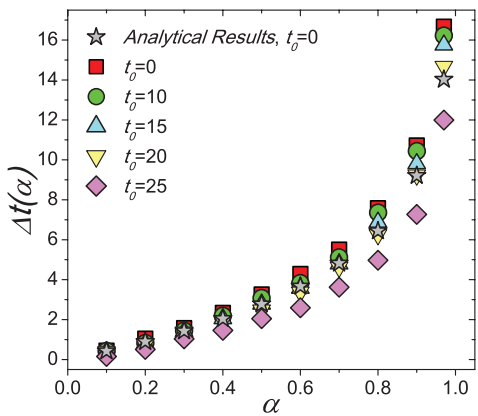}
\end{center}
\caption{(\emph{Color online}) The relation between the delay of FAT, $\Delta t(\alpha)$, and the intensity of
travel restriction. The gray stars are the analytical results with $t_0=0$. The other colored symbols are the simulation
results with various response times $t_0$=0,10,15,20, and 25 days.}
\label{fig.2}
\end{figure}

We further use the dynamic Monte Carlo method to simulate the epidemic evolution under different interventions. The simulations are
performed with discrete time steps, and we update each individual's behavior in parallel per unit time. The parameters are $N=10^6$,
$\omega=10^{-4}$, $R_0=1.75$, and $\mu^{-1}=3$ days. Initially, an infectious individual is introduced into subpopulation $x$, and thus
the initial condition is $I_{x}(0)$=1,$I_{y}(0)$=0. When the containment strategies are excluded, the epidemic reaction and diffusion
at each unit time proceed as follows. (i) Reaction: Inside each subpopulation, individuals are mixing homogeneously. At time $t$, the
probability for any susceptible in subpopulation $x(y)$ to acquire infection is $\beta I_{x}(t)/N(\beta I_{y}(t)/N)$. The number of
new infections in $x(y)$ at time $t$ is extracted from a binomial distribution with probability $\beta I_{x}(t)/N(\beta I_{y}(t)/N)$
and the number of trials $S_{x}(t)(S_y(t))$. The number of recovered individuals in $x(y)$ is also extracted from a binomial distribution
with probability $\mu$ and the number of trials $I_x(t)(I_y(t))$. (ii) Diffusion: After all individuals have been updated for the
reaction, we simulate their diffusion. The number of susceptible travelers departing from each subpopulation per unit time is also
extracted from a binomial distribution with probability $\omega$ and the number of trials $S_{x}(t)(S_{y}(t))$. The number of infectious
and recovered travelers is obtained in the same way.

We first study the effects of TR in delaying the arrival of disease to subpopulation $y$. To assemble this factor into the simulation,
we rescale the per capita diffusion rate $\omega$ by a multiplier $1-\alpha$, where the parameter $\alpha$ reflects the intensity of
TR. The strategy is activated after a given response time $t_{0}$. Figure \ref{fig.2} provides a holistic view about the relation between
the delay of FAT, $\Delta t(\alpha)$, and the restriction intensity $\alpha$. Since the disease might die out due to randomness, every
data point is obtained by averaging the simulations with the successful transfer of infectious ones among $10^4$ times of Monte Carlo random
experiments, each of which is simulated with 500 time steps. The gray stars are the analytical results obtained by Eq. (\ref{eq.3}), which
agree well with the simulations. When $\alpha=0.3,0.6$, and 0.9 and $t_0=0$, the simulations show that $\Delta t(\alpha)\simeq 2,4$, and
11 days, respectively. Even if the restriction intensity is elevated to an unpractically high level, e.g., $\alpha=0.97$, $\Delta t(\alpha)$
is still less than 3 weeks. It is clear that $\Delta t$ is small if the time scale of the initial exponential growth $1/\lambda$ is small
[see Eq. \ref{eq.3}]. We further study the impact of the response time on the efficiency of TR. In Fig. \ref{fig.2}, unless $t_{0}$ approaches
$<\!t^{F}\!>$, which is the average FAT without TR, and $\alpha$ is large, there is no evident decline for the simulation results of
$\Delta t(\alpha)$.

\begin{figure*}
\setlength{\abovecaptionskip}{-5pt}
\begin{center}
\includegraphics[width=5.5in]{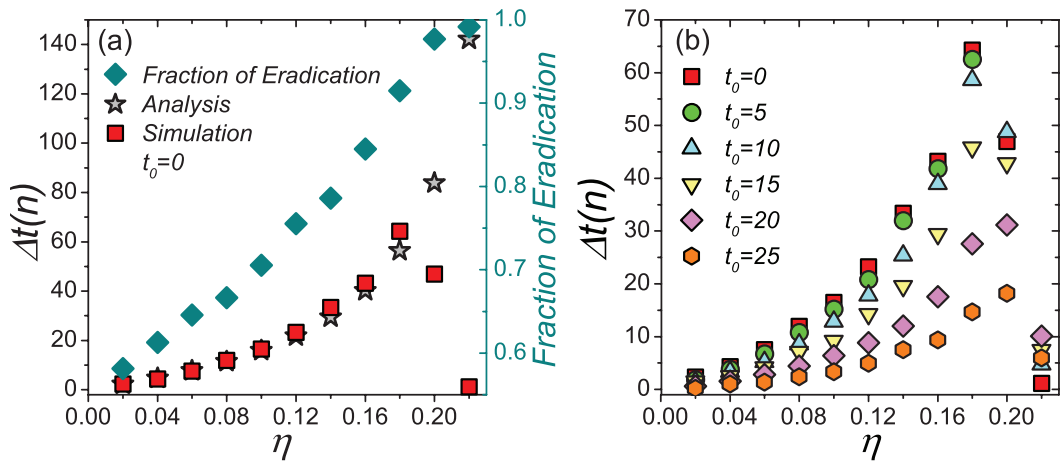}
\end{center}
\caption{(\emph{Color online}) The effects of patient isolation in delaying disease invasion. (a) The relation between the delay
of FAT$, \Delta t(\eta)$, and the isolation intensity $\eta$ with $t_{0}=0$. The dark cyan diamonds show the fraction of
eradication.(b) The simulation results with $t_{0}=0,10,15,20$, and 25.}
\label{fig.3}
\end{figure*}

We next study the effects of PI in delaying disease invasion. To introduce this factor in the model, we add an isolation process
before the reaction process at each time step after $t_0$. The parameter $\eta$ reflects the intensity of PI. Per unit time, the number
of newly isolated individuals in subpopulation $x(y)$ is extracted from a binomial distribution with probability $\eta$ and the number
of trials $I_x(t)(I_y(t))$. Figure \ref{fig.3}(a) presents the relation between the delay of FAT, $\Delta t$, and the isolation intensity
$\eta$ with $t_0=0$. For each $\eta$, we perform $10^4$ times of Monte Carlo random experiments, each of which is simulated with 500
time steps. Due to the randomness embedded in the dynamical process, the infectious individuals in source $x$ might be totally eradicated
before traveling to subpopulation $y$. With a given $\eta$, we measure $\Delta t(\eta)$ by averaging the simulations that the infectious
ones from source $x$ successfully jump to subpopulation $y$. The results are highlighted by the red squares in Fig. \ref{fig.3}(a). The
gray stars are the analytical results obtained by Eq.(\ref{eq.4}). If the isolation intensity $\eta$ is at a small or intermediate
level($\eta\le0.18$), the simulation results agree well with the theoretical predications. However, if the intensity $\eta$ is extremely
large, the simulations obviously deviate from the analytical results. In this latter case, since the Malthusian parameter $\lambda_{\eta}$
is quite small, there is a huge likelihood of totally eradicating the infectious individuals at the early stage of an outbreak due to
randomness. For instance, when $\eta=0.2,0.22$, the fraction of eradication in all independent modeling realizations reaches 97.7$\%$
and 99.2$\%$, respectively, while for $\eta=0.12$, the fraction of eradication is only 75.6$\%$ [see the dark cyan diamonds in
Fig.\ref{fig.3}(a)]. If $\eta\ge0.25$, the Malthusian parameter $\lambda\leq 0$, the disease hardly persists in the population. With the
same condition that $t_0=0$, the strategy of PI is more efficient than TR: An intermediate level of isolation intensity $\eta$ can
adequately delay the arrival of disease for about 1 month.

Figure \ref{fig.3}(b) shows the impact of the response time $t_0$ on the delaying effects of PI. For a small $t_0$, e.g., $t_0=10$,
which is much smaller than $<\!t^F\!>$, an intermediate level of PI(e.g., $\eta=0.14$) still suspends the arrival of disease for
about 3 weeks. This achievement exceeds the performance of TR even with an extremely high restriction intensity. The simulations
also illuminate that the PI is sensitive to the increase of $t_0$. There is a remarkable decline in the simulation results of
$\Delta t(\eta)$ when $t_0$ approaches $<\!t^F\!>$. For instance, 25 days of waiting to implement the strategy ($t_{0}=25$) will
only postpone the arrival of disease in subpopulation $y$ for about 2 weeks at most.

Actually, other intra-population interventions can also be analyzed under this framework. For instance, social distancing limits
public activities to reduce personal contacts, which can be reflected by rescaling the disease transmission rate $\beta$ with a
multiplier $1-\varphi$ when time $t\ge t_0$. At the initial stage of an outbreak, the Malthusian parameter becomes
$\lambda_{\varphi}\!=\!(1-\varphi)\beta-\mu$. From a mathematical point of view, we can adjust the parameters $\varphi,\eta$ to allow $\lambda_\varphi\!=\!\lambda_{\eta}$. Therefore, the above analysis can cover this scenario.

\section{SUMMARY}

In sum, the intra-population interventions, e.g., patient isolation, perform better than the inter-population strategies such as
travel restriction if the response time is small. Therefore, the intra-population strategies are more beneficial in delaying
the spatial spread of pandemic influenza if they are implemented very promptly. However, the intra-population measures are sensitive
to the increase of response time, which might be inevitable due to miscellaneous socioeconomic reasons in reality and largely
discounts the efficiency.

\section{ACKNOWLEDGMENTS}

This study is supported by National Key Basic Research and Development Program (No.2010CB731403), the NCET program (No.NCET-09-0317),
and the National Natural Science Foundation(No. 61273223) of China.


\begin{thebibliography}{57}



\bibitem{ANDERSONMAY}
R.M. Anderson, R.M. May, \emph{Infectious Diseases of Humans: Dynamics and Control} (Oxford University Press, Oxford, U.K., 1991).



\bibitem{PR424175}
S. Boccaletti, et al., Phys. Rep. {\bf 424}, 175(2006).


\bibitem{BBV08}
A. Barrat, et al. \emph{Dynamical Processes on Complex Networks}(Cambridge University Press, Cambridge, U.K., 2008).

\bibitem{RMP801275}
S.N. Dorogovtsev, et al., Rev. Mod. Phys. {\bf 80}, 1275(2008).


\bibitem{NP832}
A. Vespignani, Nat. Phys. {\bf 8}, 32(2012).



\bibitem{PRE63066117}
R. Pastor-Satorras and A. Vespignani, Phys. Rev. Lett. {\bf 86}, 3200(2001);
Y. Moreno, et al., Eur. Phys. J B {\bf 26}, 521(2002); Bogu\~{n}\'{a} M., et al.,
Phys. Rev. Lett. {\bf 90}, 028701(2003); Xia C.Y., et al., Int. J Mod. Phys. B
{\bf 23}, 2303(2009); B. Guerra, J. G\'{o}mez-Garde\~{n}es, Phys. Rev. E {\bf 82},
035101(2010); C. Castellano and R. Pastor-Satorras. et al., Phys. Rev. Lett. {\bf 105}, 218701(2010).




\bibitem{MATHBIO753}
L.A. Rvachev, and I.M. Longini, Jr., Math. Biosci. {\bf 75}, 3(1985);
C. Viboud, et al., Science {\bf 312}, 447(2006);
V. Colizza, et al., Nat. Phys. {\bf 3}, 276(2007);
D. Balcan, et al., Proc. Natl. Acad. Sci. U.S.A. {\bf 106}, 21484(2009);
Wang L., et al., PLoS ONE {\bf 6}, e21197(2011).
H.H.K Lentz, et al., Phys. Rev. E {\bf 85}, 066111(2012).
C. Poletto, et al., Sci. Rep. {\bf 2}, 476(2012).

\bibitem{PRE84041936}
L. Cao, et al., Phys. Rev. E {\bf 84}, 041936(2011)





\bibitem{WHO}
World Health Organization, \emph{Pandemic Influenza Preparedness and Response (WHO, Geneva, 2009)};
United States Department of Health and Human Services, \emph{HHS Pandemic Influenza Plan (HSS, Washington, D.C., 2005)}.






\bibitem{SCI326729}
Y. Yang, et al., Science {\bf 326}, 729(2009);
J. G\'{o}mez-Garde\~{n}es, et al., Proc. Natl. Acad. Sci. U.S.A. {\bf 105}, 1399(2008); 
P. Holme, Europhys. Lett. {\bf 68}, 908(2004);
M.E. Halloran, et al., Proc. Natl. Acad. Sci. U.S.A.S {\bf 105}, 4639(2008);
Y. Chen, et al., Phys. Rev. Lett. {\bf 101}, 058701(2008);
L. Hufnagel, et al., Proc. Natl. Acad. Sci. U.S.A. {\bf 101}, 15124(2004);
V. Colizza, et al., PLoS Med. {\bf 4}, e13(2007);


\bibitem{SCI3091083}
I.M. Longini Jr., et al., Science {\bf 309}, 1083(2005).


\bibitem{PLOSMED3e212}
B.S. Cooper, et al., PLoS Med. {\bf 3}, e212(2006).


\bibitem{NatMed12497}
T.D. Hollingsworth, et al., Nat. Med. {\bf 12}, 497(2006).



\bibitem{PLoSONE2e401}
J.M. Epstein, et al., PLoS ONE {\bf 2}, e401(2007).


\bibitem{PLoSONE6e16591}
P. Bajardi, et al., PLoS ONE {\bf 6}, e16591(2011).



\bibitem{MB21470}
G.S. Tomba, J. Wallinga, Math. Biosci. {\bf 214}, 70(2008).

\bibitem{PLoSONE2e143}
P. Caley, N.G. Becker, D.J. Philp, PLoS ONE {\bf 2}: e143(2007).


\bibitem{JSM09001}
A. Gautreau, et al., J. Stat. Mech. L09001(2007).

\bibitem{JTB251509}
A. Gautreau, et al., J. Theor. Biol. {\bf 251}, 509(2008).

\bibitem{PNAS1068847}
A. Gautreau, et al., Natl. Acad. Sci. U.S.A. {\bf 106}, 8847(2009).

\bibitem{PRL103038702}
J.L. Iribarren, E. Moro, Phys. Rev. Lett. {\bf 103}, 038702(2009).

\bibitem{Interface7873}
O. Diekmann, et al., J. R. Soc. Interface {\bf 7}, 873(2010).

\end{thebibliography}
\end{document}